\documentclass[aps,preprint,superscriptaddress,showpacs]{revtex4}
\usepackage{amsmath,bm}
\usepackage{graphicx}

\begin{document}

\title{Modification of the electronic structure in a carbon nanotube with the charge dopant encapsulation}
\author{Woon Ih Choi}
\affiliation{Department of Physics and Astronomy, Seoul National
University, Seoul 151-747, Korea}
\author{Jisoon Ihm}
\affiliation{Department of Physics and Astronomy, Seoul National
University, Seoul 151-747, Korea}
\author{Gunn Kim}
\email[corresponding author. Electronic address:\
]{kimgunn@skku.ac.kr} 
\affiliation{BK21 Division and Department of
Physics, Sungkyunkwan University, Suwon, 440-746, Korea}
\date{\today }

\begin{abstract}

We present the first-principles study of effects of the
charge dopants such as Cesium and Iodine encapsulated on the
electronic structure of carbon nanotubes. An encapsulated cesium
atom donates an electron to the nanotube and produces donor-like states
below the conduction bands. In contrast, an iodine trimer
(I$_{3}$) accepts an electron from the nanotube and produces an
acceptor-like state above the valance band maximum. 
We find that a Cs atom inside a metallic armchair carbon nanotube gives rise to
spatial oscillations of the density of states near the Fermi
level.

\end{abstract}

\maketitle

The tubular form which has the hollow space
inside makes researchers insert atoms or molecules such as
fullerenes\cite{Takenobu-NatureM, Lee-Nature, Hornbaker-Science, Cho-PRL} 
and hydrocarbons\cite{Kim-CPL,Steinmetz-CPL,Steinmetz-CAP} to obtain useful physical properties.
The doping of a semiconductor is essential to be used for electronic devices. 
Recent experiments have reported that encapsulation of alkali metal
(Cs) or halogen elements (I) inside the carbon nanotube (CNT) 
is feasible.\cite{Jeong-PRB, Bendiab-PRB, double-helix-I-PRL-2000, P-dependence-ICNT-PRB-2002}  
It has been reported that the
field emission and electrical current can be improved by doping of these
materials.\cite{Cscnt-FE-enhance-APL-2006, Icnt-elec-apl-2007}
The scanning tunneling spectroscopy (STS) study of 
the Cs-encapsulated CNT shows local change of electronic structure.\cite{Kim-PRL} 
The previous theoretical studies, however,
considered only a high doping (dense packing of dopants).
The effect of the charge donor or acceptor on the local electronic
structure of semiconducting or metallic CNT without the
interaction between the neighboring images in the periodic
supercell has not been fully understood yet.

In this Letter, we investigate the influence of an encapsulated charge doping atom (Cs, I) on the
electronic structure of CNTs, using the first-principles density
functional calculations. For the Cs-encapsulated CNT,
an electron is transferred from an inserted Cs atom to the CNT and
produces donor-like states below the conduction bands. For the
I$_{3}$-encapsulated CNT, in contrast, the iodine trimer (I$_{3}$)
accepts an electron from the CNT and produces an acceptor-like
state above the valance band maximum. 
These findings may lead us to understand the local electronic properties of dopant-encapsulated CNTs.


We performed ab initio density
functional calculations within the local density approximation
using a pseudo-atomic orbital basis set.\cite{Ozaki-PRB1, Ozaki-PRB2} 
The norm-conserving pseudopotential
\cite{Hamann-Schluter-Chiang} 
is employed with an energy cutoff of 120 Ry. We adopt Ceperley-Alder exchange-correlation potential.\cite{Ceperley-PRL}
Here we choose Cs and I as an electron donor and
an acceptor, respectively. The scalar relativistic effects important
for heavy atoms such as Cs and I are taken into account. 
We treat 4 electrons (2s$^{2}$2p$^{2}$) for C, 9 electrons (5s$^{2}$5p$^{6}$6s$^{1}$) for Cs and 17 electrons (4d$^{10}$5s$^{2}$5p$^{5}$) for I as valence electrons.
 For dopant encapsulation, the semiconducting
(10, 0) and metallic (5, 5) CNTs which have similar diameters ($\sim$ 7 \AA)
are chosen. The periodicity of CNTs along the tube axis is from 8
nm to 10 nm, containing 800 - 1000 carbon atoms. The
I$_{3}^{-}$ or I$_{5}^{-}$ are experimentally found to form
\cite{Bendiab-PRB} inside the CNT, which was confirmed by the
first-principles study. Therefore, we put an iodine atom or a trimer (I$_{3}$)
inside the CNT. The energy minimum position of the encapsulated
charge dopant atom is found to be at the axis of the nanotube.


To see the spatial change of the electronic structure, 
we constructed the simulated STS data with the projected density of states (PDOS) of carbon
atoms which lie in different positions along the CNT axis. At
first, the optimized geometry of the Cs-encapsulated (10, 0) CNT
and the simulated STS map of the structure are presented in Fig.
1(a). A violet-colored sphere at the center of the CNT represents a
Cs atom. We found that the electronic structure of the CNT near
the Cs atom is changed and it recovers the original DOS of the
(10, 0) CNT far from the Cs site. Our Mulliken population analysis
shows that the Cs atom gives approximately one electron to the CNT
and becomes a Cs$^{+}$ ion. The carbon $\pi$ electrons which were
originally extended over the whole nanotube become bound near the
Cs$^{+}$ ion. The local electronic states are overall downward
shifted and bound states are produced near the Cs-embedded
site. In Fig. 1(a), one of the bound states is shown
in the energy band gap. According to the
angular momentum ($l = 3$) of the bound states, this state originates
from the conduction band minimum (CBM; $\sim$ 0.25 eV). The other
two bound states occur ($\sim$ 0.5 and $\sim$ 0.7 eV) above the Fermi level.
Besides relatively localized states with high densites are shown above the 1.5 eV. 
We observe the DOS depression of the CBM near the Cs atom. The DOS of
the valence band also decreases because of the local downshift.
These results are in good agreement with experiment.\cite{Kim-PRL}

Next, we investigate the hole doping effect of iodine. Recent
experiment reports that iodine atoms inside the CNT form I$_{3}$
or I$_{5}$. The short chain structures accept an elelctron
from the nanotube and become I$_{3}^{-}$ or I$_{5}^{-}$.\cite{Bendiab-PRB} The electronic structure and geometry of
I$_{3}$-encapsulated CNT is shown in Fig. 1(b). An acceptor-like
state bound near the iodine trimer appears near the
valence band maximum (VBM). The depression of the DOS of the valence band is observed
near the I$_{3}$ position since the acceptor-like state originates
from the VBM. As expected, this tendency is the opposite of
the case where a Cs atom is encapsulated inside the tube. 
The transferred electrons are not uniformly distributed on the carbon nanotube. 
Like the local densities of states (LDOS) in Figs. 1(a) and 1(b), 
according to the Mulliken population anaysis, 
the total electron density along the tube axis has a decay pattern, which means the local electron transfer. 
Regardless of the structural optimization (relaxation), this feature is shown. 
We conclude that local mechanical deformation of the CNT is not very important 
to the modulation of the LDOS.

We also consider the case where two different kinds of charge dopants are encapsulated together
inside the CNT. We put Cs and I atoms with the distance of 4 nm. 
As shown in Fig. 2 (a), donor-like and acceptor-like states are
shown near Cs and I embedded sites, respectively. The several
horizontal cuts of Fig. 2 (a) are shown in Fig. 2 (b). The smooth
curves are shown for visual clarity. The potential
from Cs$^{+}$ and I$^{-}$ ions makes the VBM and CBM oscillate like
sinusoidal waves with the size of the band gap preserved. 
It is clearly shown that the local shift of the potential in the cesium site is opposite to 
that in the iodine site.
In addition to the bound states in the gap region, we can
see the quantum well states by the dopant ion in both of valence and conduction bands. The confined states in Fig 2 (a) are
located at the upper left and lower right region. The number of nodes in those states increases as energy
increases.

Finally, we investigate the effect of encapsulated
charge dopants on the metallic armchair CNT. We select an
armchair (5, 5) CNT which has robust metallic properties by the
symmetry. In the simulated STS image shown in Fig. 3 (a), there are 
states ($-$3.5, $-$2.2, and $+$1 eV) which have localization characters. 
To see the detailed spatial change of the DOS near the Fermi level, the enlarged picture of
Fig. 3 (a) is shown in Fig. 3 (b). 
Considering horizontal cuts in Fig. 3, the spatial oscillation of the
DOS are shown. We find that the screening amplitude and
frequency are both dependent on energy. The positive charge (Cs$^{+}$) has the screening length of $\sim$0.5 
nm and there are the residual screening patterns over the radius.
It demonstrates that the free-electron screening
length in one-dimensional metallic structures is longer than in
three-dimensional metals.\cite{Feng-PRL}

By the Fourier transform for the spatial oscillation patterns, some features are found as follows: 
In k-space, all peaks in the Fourier-transformed patterns map onto linear $\pi$ and $\pi^{*}$ bands of the
armchair CNT near the Fermi level. Below the Fermi level,
both of two linear $\pi$ and $\pi^{*}$ bands contribute to the screening. 
Above the Fermi level, however, only the $\pi$ band contributes to the screening of the Cs$^{+}$ ion. 
We note that the wave functions of the $\pi$ states do not have phase variation
in the circumferential direction in the armchair CNT while the
wave functions of $\pi^{*}$ electrons are more complicated to
match and to interact with the orbitals of Cs because of the
phase variation along the circumferential direction.
This phase variation of $\pi^{*}$ states in the circumferential direction is prominent in the unfilled (conduction) band.


In summary, effects of charge doping on the electronic structure
of CNT have been investigated with first-principles calculations.
Halogen and alkali metal atom produce donor-like and
acceptor-like states in the gap, respectively. Our work shows that
the charge transfer by the dopant\cite{Smally-chTransfer-nature-1997} 
and the generation of gap states greatly affect the electronic structure of the CNT.
In addition to the bound states in the gap region, we find 
the quantum confined states by the dopant ion in both of valence and conduction bands. 
As in quantum harmonic oscillators, the number of nodes in those electronic states increases as energy
increases in the conduction band.
Dopant-encapsulated metallic armchair CNTs show the spatial
oscillations of its density of states near the Fermi level which
is related to the linear bands crossing at the Fermi level.


We are grateful to S.B. Lee and J. Im for fruitful discussions.
This work is supported by the STAR-faculty project of Ministry of Education, 
the SRC program (Center for Nanotubes and Nanostructured Composites) of MOST/KOSEF and
the second BK21 project. Computations are performed through the support of KISTI.

\newpage
\begin{center}
\LARGE{[Figure Captions]}
\end{center}

Figure 1: (online color) (a) Two-dimensional map of simulated STS of a (10, 0)
CNT encapsulating a Cs atom and its model
structure. (b) Two-dimensional map of simulated STS of a (10, 0)
CNT encapsulating a iodine trimer (I$_3$) and its model structure.
The Fermi level is set to zero. The orange and brown spheres located
in the middle of CNT represent Cesium and Iodine, respectively.

Figure 2: (online color) (a) Two-dimensional map of simulated STS data
of a (10, 0) CNT encapsulating Cs and I and its model structure. 
The horizontal cuts of the STS map in several energies are shown in (b). The curves are shown as smooth curve again for visual clarity.

Figure 3: (online color) Two-dimensional map of simulated STS data of a (5, 5) CNT encapsulating a Cs atom and its model structure are
shown in (a). The enlarged STS map is shown again in (b) to see the detailed structure of the DOS near the Fermi level.

\newpage

\newpage
\begin{figure}[t]
  \centering
  \includegraphics[width=10.0cm]{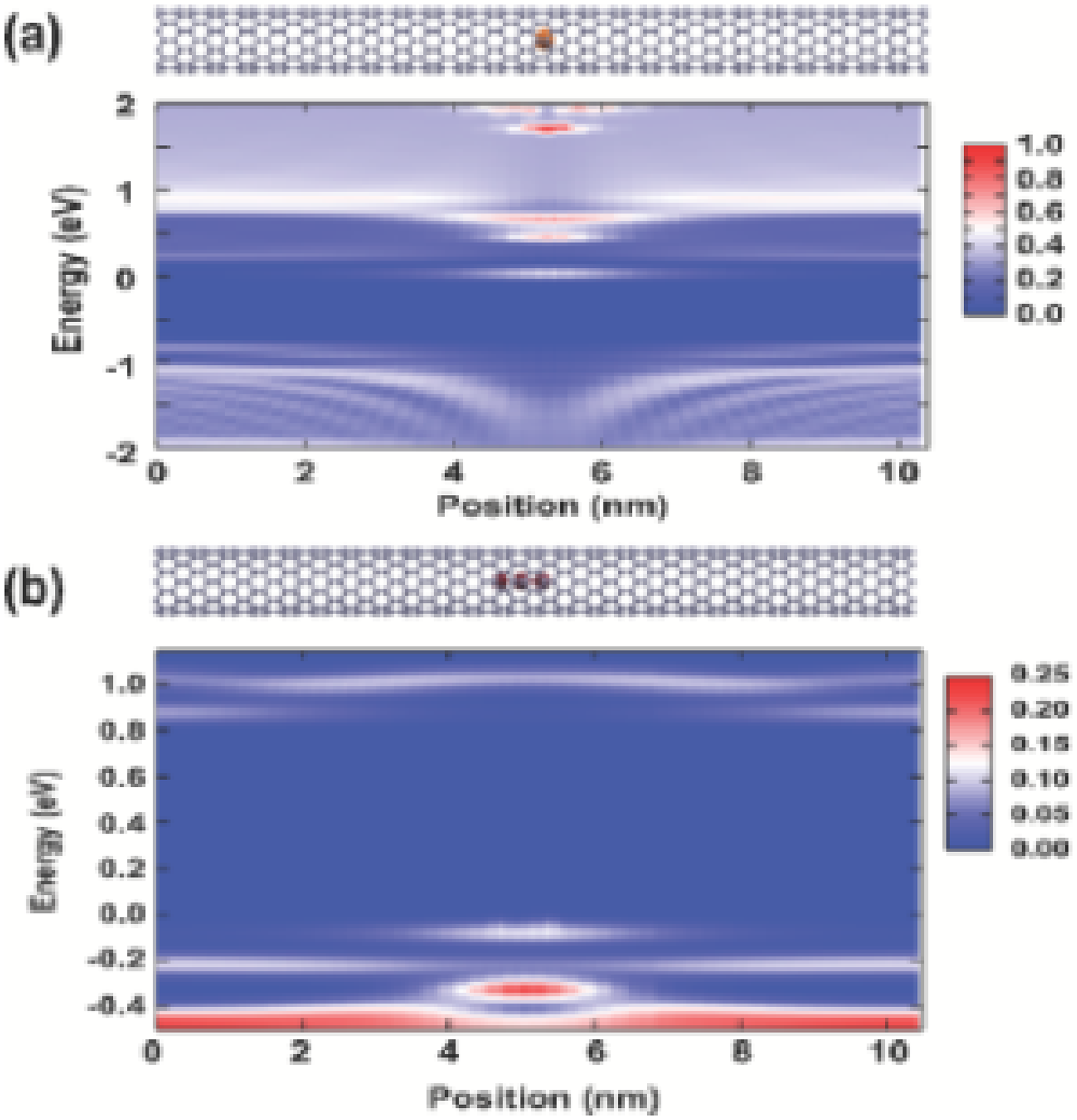}
\end{figure}
\begin{center}
\LARGE{Figure 1}

\LARGE{W.I Choi et al.}
\end{center}

\newpage
\begin{figure}[t]
  \centering
  \includegraphics[width=10.0cm]{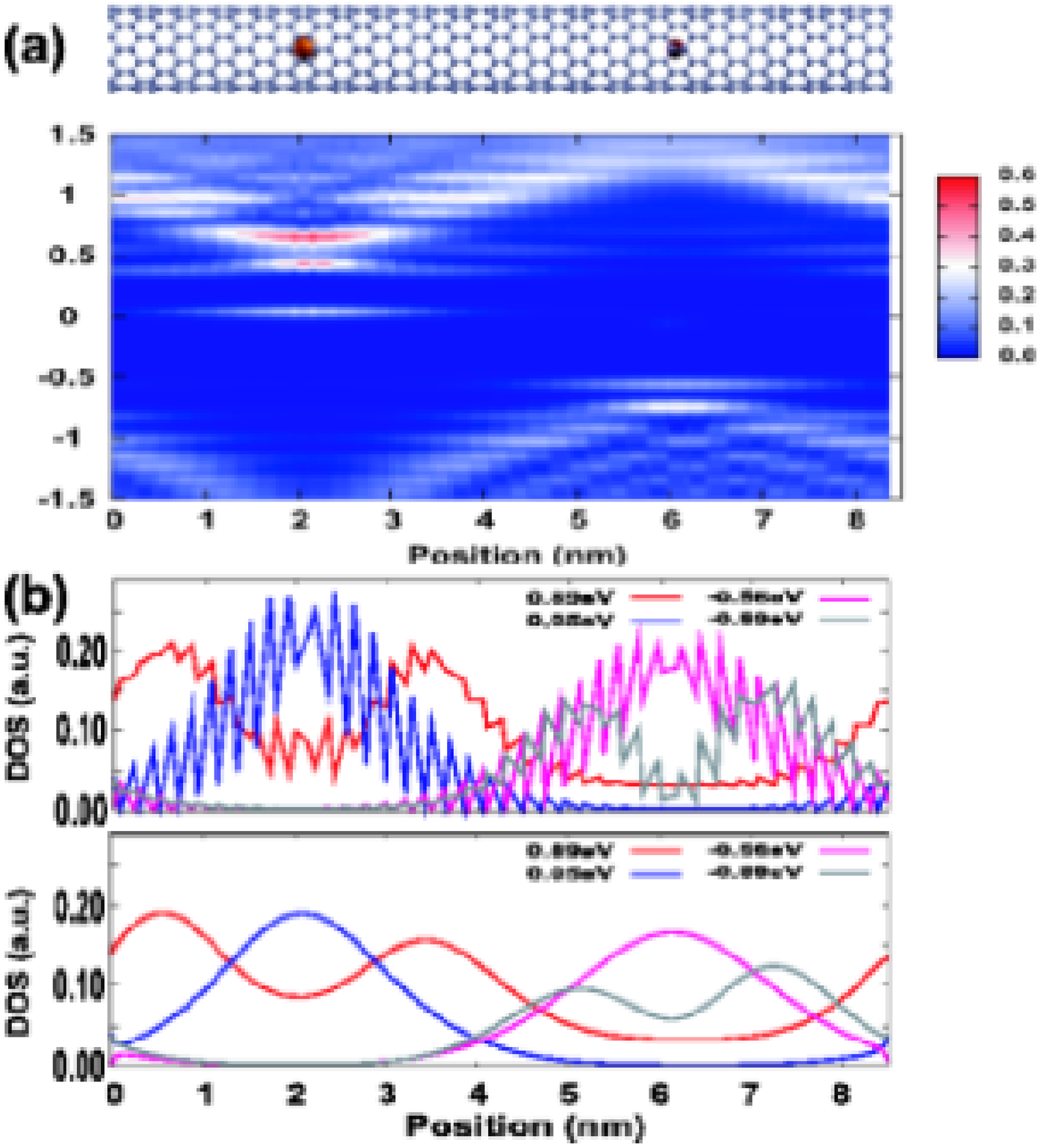}
\end{figure}
\begin{center}
\LARGE{Figure 2}

\LARGE{W.I. Choi et al.}
\end{center}

\newpage
\begin{figure}[t]
  \centering
  \includegraphics[width=10.0cm]{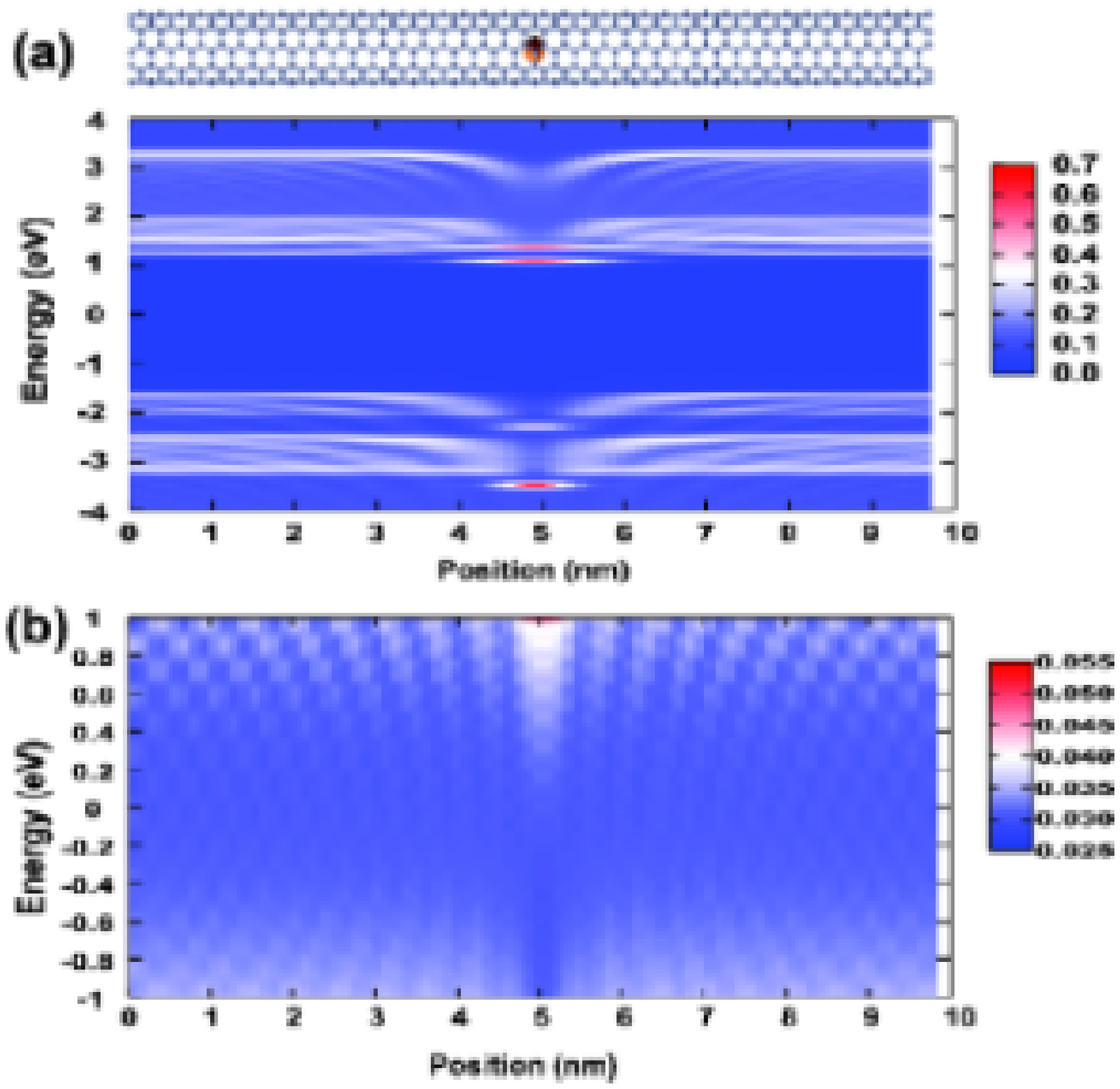}
\end{figure}
\begin{center}
\LARGE{Figure 3}

\LARGE{W.I. Choi et al.}
\end{center}

\end{document}